\documentclass[]{spie}  

\usepackage[]{graphicx}
\def\titlefont{\LARGE\bf} 
\title{Quantum stream cipher based on optical communications} 

\author{
Osamu Hirota\supit{a,b}, 
Kentaro Kato\supit{b}, 
Masaki Sohma\supit{a},
Tsuyoshi S. Usuda\supit{c},\\
Katsuyoshi Harasawa\supit{d}
\skiplinehalf
\supit{a}
Research Center for Quantum Information Science, Tamagawa University, 
 Tokyo, Japan\\
\supit{b}
21st century COE program, Chuo University, Tokyo, Japan\\
\supit{c}
Aichi Prefectural University, Aichi, Japan\\
\supit{d}
Hitachi Hybrid Network Co. Ltd. Yokohama, Japan.
}

\authorinfo{
Further author information:\\
Osamu Hirota: address; 6-1-1, tamagawa-gakuen, machida-city, Tokyo, Japan. E-mail, hirota@lab.tamagawa.ac.jp, Telephone, +81-42-739-8674.\\
Kentaro Kato:E-mail, kkatop@ieee.org. Masaki Sohma:E-mail, sohma@eng.tamagawa.ac.jp. Tsuyoshi S.Usuda:E-mail,usuda@ist.aichi-pu.ac.jp
}

\begin{document} 
  \maketitle 

\begin{abstract}
In 2000, an attractive new quantum cryptography was discovered by H.P.Yuen based on quantum communication theory. It is applicable to direct encryption, for example quantum stream cipher based on Yuen protocol(Y-00),  with high speeds and for long distance by sophisticated optical devices which can work under the average photon number per signal light pulse:$<n> = 1000 \sim 10000$. In addition, it may provide information-theoretic security against known/chosen plaintext attack, which has no classical analogue.
That is, one can provide secure communication, even the system has $H(K) << H(X)$. 

In this paper, first, we give a brief review on the general logic of Yuen's theory. Then, we show concrete security analysis of quantum stream cipher to quantum individual  measurement attacks. Especially by showing the analysis of Lo-Ko known plaintext attack, the feature of Y-00 is clarified. In addition, we  give a simple experimental result on the advantage distillation by scheme consisting of intensity modulation/direct detection optical communication.
\end{abstract}


\section{INTRODUCTION}
There is no encryption scheme with provable security in the conventional cryptography. One of methods to provide "provable security" is quantum cryptography.
A quantum key generation scheme  for two legitimate users(Alice and Bob) as the quantum cryptography is one of the most interesting subjects
in quantum information science, which was pioneered by C.Bennett and G.Brassard in 1984[1]. We emphasize that such results are great achievement and open a new science.  
Many researchers believe that the key distribution by single photon is on the verge of commercial application. However we should take into account the fact that the societies of electronics and communication, and of cryptography are basically not interested in the practical use of quantum cryptography based on single photon schemes, because of extremely low performance in the sense of communication science. Although there is no means of solving such a serious argument,
we would like to  make the following comment.  The key generation is a very important, but it is very narrow sense that one defines quantum cryptography by  only  BB-84 and  similar principle[2].
In addition, still it requires "one time pad" to provide secure communication in principle.

Yuen, and his group have pointed out that the quantum cryptography should involve other aspects, and called  quantum information scientist's attention to quantum cryptography based on another principle[3][4]. The basic protocol is called Yuen protocol 2000(Y-00).
The fundamental structure of Y-00 is organized by a shared initial key like symmetric key scheme in the conventional cryptography, but it is constructed as a physical cryptography. In addition, this gives a generalization of conventional unconditionally secure key generation based on Maurer[5] and similar theory in which a shared seed key is not used to establish an advantage distillation. The research like Gisin's work[6] and coherent state based BB-84[7] to cope with the low performance should be encouraged, and also  research like Northwestern University's group to investigate another scheme[8,9] for achieving the same function should be welcome. The authors believe that collaboration of both types of the quantum cryptography brings real applications in communication networks, because the combination of BB-84 and AES has no meaning in the sense of secure communication.

In this paper, to reveal an excellent  potential of Y-00,  we shall discuss on concrete security analysis of quantum stream cipher as an example of Y-00. The paper is organized as follows. We introduce the new quantum cryptography  in the section 2 and give concrete performance of quantum stream cipher in the section 3. In addition, an experimental result of quantum stream cipher based on intensity modulation and direct detection scheme as conventional optical communication is reported.

\section{Yuen protocol:Y-00}
In this section, we will survey a theory of Y-00[4].
First we assume that Alice and Bob share a seed key $K$. The key is stretched by a pseudo random number generator to $K'$.
The data bit is modulated by $M$-ary keying driven by random decimal number generated from the block :$K'/log M=\bar{K}'=(k_1,k_2,\dots)$ of pseudo random number with the seed key $K$. 
The $M$-ary keying has $M$ different basis based on 2$M$ coherent states. So the data bit is mapped into one of 2$M$ coherent states randomly, but of course its modulation map has a definite relationship based on key, which is opened.

Let us  mention first what is  basic principle to guarantee the security.
There are many  fundamental theorems in quantum information theory. The most important theorem for information processing of classical information by quantum states is the following: 
\\
{\bf Theorem 1}: \\
{\it Signals with non-orthogonal states cannot be 
distinguished without error and  optimum lower bounds for error rate exist.}\\
This means that if we assign non-orthogonal states for bit values 1 and 0,
then one cannot distinguish 1 and 0 without error.
When the error  probability is 1/2 based on quantum noise, there is no way to distinguish them, from quantum signal detection theory pioneered by Helstrom, Holevo, and Yuen[10].
On the other hand, there is no-cloning theorem developed by Wootters-Zurek, and Yuen[11] as follows.\\
{\bf Theorem 2}: \\
{\it Non-orthogonal states cannot be cloned without error.}\\
This is an essential basis for BB-84 and others. However Y-00 does not require this theorem explicitly .

\subsection{Direct data encryption}
An application of Y-00 is, first,  direct data encryption like a stream cipher in conventional cryptography, and then it is extended to key generation, but  not  one-time pad which is very inefficient. Here it is natural that we should employ different security criteria for  direct encryption and  key generation.
For  direct encryption, the criteria are given as follows.
\begin{itemize}
\item[\rm(i)]  Ciphertext-only attack on data and on key: To get plaintext or key, Eve knows only the ciphertext from her measurement.
\item[\rm(ii)] Known/chosen plaintext attack: To get key, Eve  inserts  her known or chosen plaintext data into modulation system( for example, inserts all 0 sequence as text). 
Then Eve tries to determine key from input-output. Using the key, Eve can determine the data from the ciphertext.
\end{itemize}

In the conventional theory, we have $H(X|Y_E, R_M) \le H(K)$  known as Shannon bound for ciphertext only attack on data, and $H(K|Y_E,R_M)\ge 0$ for ciphertext only attack on key which is relevant with "unicity distance", where $X$ is data sequence, $Y_E$ is Eve's data, and $R_M$ is public mathematical randomization, respectively. In addition, for known plaintext attack, we have $H(K|X, Y_E, R_M) =0$ which means a computational complexity based security.
In the following, we will see that one may overcome such limitations in the conventional direct encryption by Y-00.

A fundamental requirement of secure communication by "one way scheme" is, first,
to establish  that the channel between Alice and Eve is very noisy, but the channel of Alice and Bob is kept as a normal communication channel by physical structure. To realize it, a combination of  a  shared short  key  for the legitimate users and a kind of stream cipher with specific modulation scheme is employed, following the theorem 1. We note that a main idea of this protocol is the explicit use of a shared short key and physical nature of scheme for cryptographic objective of secure communication and key generation. This is called initial seed key advantage. \\

{\bf Principle of security}: 
{\it The origin of security comes from difference between optimum quantum measurement performances with key and without key.}\\

In general, a quantum information system is described by a density operator. The density operator of the output of the coding/modulation system of Y-00 for Eve depends on attacks. 
For ciphertext-only quantum individual attack, the density operator is
\begin{equation}
\rho_{T}=p_0\rho_0 + p_1\rho_1
\end{equation}
where
\begin{equation}
\rho_0 = \sum q_j|\alpha_j\rangle\langle \alpha_j|, \quad 
\rho_1 = \sum q_k|\alpha_k\rangle\langle \alpha_k|
\end{equation}
The probability $p_i$ depends on the statistics of the data, and $q_j$, $q_k$ depend on the pseudo random number with $j$, and  $k$ being even and odd number, for example. Eve has to extract the  data from the quantum system with such a density operator.  However, according to  one of the most fundamental theorem(theorem 1) in quantum information theory, the accuracy of Eve's measurement is limited.
Thus, to induce error in Eve's measurement is essential in Y-00.
Error probability of Eve and its requirement for secure system depend on attack methods.
A typical measurement in the quantum individual attack is  direct measurement of transmitting signal, which corresponds to discrimination of information bit or 
discrimination of $M$-ary states. In this case, for  ciphertext only attack on the information data, the best way of Eve is of course given by the quantum optimum detection for two mixed states :$\rho_0$ and $\rho_1$. That is, the limitation for accuracy of measurement of Eve is given by Helstrom bound as follows[10]:
\begin{equation}
\bar{P}_e= \min_{\Pi}(p_1Tr\rho_1\Pi_0 + p_0Tr\rho_0\Pi_1)
\end{equation}
where $\Pi$ is POVM(positive operator valued measure) or quantum detection operator which corresponds to general optical receiver in optical communications. The error probability of Eve becomes $\sim 1/2$ from the appropriate choice of the number $M$ and signal energy. It means that Eve's data $Y_E$ is completely inaccurate.
On the ciphertext only attack on key, the best way for Eve is to detect $M$ basis based on 2$M$ coherent states. In this case, the limitation for accuracy of Eve's data is also given by the minimax quantum detection[12] for 2$M$ pure coherent states. That is, the measured data on the running key involve unavoidable error given by 
\begin{equation}
\bar{P}_e = \max_{p_i}\min_{\Pi} (1 - \sum p_iTr \rho_i\Pi_i)
\end{equation}
For appropriate $M$ and signal energy, we have $\bar{P}_e \sim 1$. 
As a result, Eve's data have almost complete error by irreducible noise. 

When Eve can know or insert some input data $X$, one can apply also the above equation, but the number for detection is reduced to $M$ coherent states. On the other hand, Eve can devise certain known/chosen plaintext attack in this situation. That is, Eve can use the known plaintext after all trial of her measurement.

If Eve takes another type of attack like combination with measurement of {\it indirect observable } and structure information of $M$-ary modulator, then  density operators of Eve may be redefined and the error probability is also calculated by quantum detection theory as shown in later sections.

Let us turn to issue of randomization which is essential in Y-00 for getting ultimate performance in practical sense. The structure of Y-00 is formed by physical processes with specific modulator performance and so on. This means that the protocol is different with conventional cryptography formulated by mathematical relation only, even the scheme is constructed by devices based on classical physics. It  is called "physical cryptography". Since the protocol is constructed by combination of physical processes and mathematical encryption, one can devise a new randomization that is possible with physics to increase the security. Such a randomization $R_P$ is called "physical randomization". A general theory of such randomizations was already given, which involves a theory of  {\bf DSR}: deliberate signal randomization, {\bf DER}:deliberate error randomization and so on [4]. Thus, this is a new type of quantum cryptography based on quantum detection theory. That is, the security is guaranteed by quantum noise, but the system can be implemented by sophisticated optical devices or quantum optimum receiver, depending on the required performance. As an advantage,  Y-00 may provide information-theoretic security  against  known/chosen plaintext attack based on several quantum measurements. 

We can summarize the properties as follows:\\
In the cipher-text only attack on data, Y-00 can exceed the classical Shannon limit, even we use a system with $H(K) << H(X)$. That is,
\begin{equation}
H(X|Y_E, R_M, R_P) > H(K)
\end{equation}
where $X$ is information data, $Y_E$ is ciphertext as "measured value" for Eve, $R_P$ is physical randomization and $K$ is initial seed key. 

For known/chosen plaintext attack, it will be expected that 
\begin{equation}
H(K|Y_E,R_M, R_P, X) > 0
\end{equation}
which corresponds to {\bf information-theoretic security}. If one has $H(K|Y_E,R_M, R_P, X) =H(K)$, then it is full security. 
These are not realized by the conventional theory. It means that Y-00 breaks the limitation of the conventional cryptography theory.
Furthermore, by introducing several new randomizations, this will  provide information-theoretically secure scheme with highly efficient performance  even if the advantage of the legitimate user is small in principle.

When the signal power is very strong and the system has a bad design, there may be some algorithm for computing to find the key based on the measured data including "small error". It may have $H(K|Y_E, R_M, R_P, X) \sim 0$. Even so, it requires additional exponential search.

\subsection{Quantum key generation}
In the case of key generation protocol, data is a true random number sequence. So there is no criterion like known plaintext attack. In the conventional theory of key generation, if one has 
\begin{equation}
I(X_A; Y_E) < I(X_A; Y_B),
\end{equation}
then key generation is possible.
That is, an information-theoretic existence proof is given under "the condition of no public discussion". However, so far there was no theoretical consideration on scheme with shared seed key between Alice and Bob. 
In the above subsection, we explained the basic scheme. By this scheme or more generalized scheme, one can also make a key generation scheme. The use of shared seed key between Alice and Bob that determine the quantum states generated for the data bit sequences in a detection/coding scheme gives them a better error performance  over Eve who does not know $K$.
Based on the above scheme, the conditions for key generation were discussed to include the use of {\it a shared seed key}. Let us introduce the basic result here.\\
{\bf Remark}\\
{\it For the scheme with the seed key, one has to make sure that the generated key:$K_g$ between Alice and Bob is fresh. That is, the generated key is independent of the initial seed key, and} $I(K_g; Y_E |K) \sim 0$.

In order to characterize this situation, one can allow that 
Eve can know  the key only after she has made her measurement.
So her information is described by $I(X_A, Y_E|K)$.
As a result, the condition for secure key generation is 
\begin{equation}
I(X_A; Y_E|K) < I(X_A; Y_B)
\end{equation}
or 
\begin{equation}
H(X_A|Y_E, K) > H(X_A|Y_B)
\end{equation}
where $Y_B$ is Bob's observation with knowledge of the seed key. 
The above result means that Eve can get the key $K$ after Eve's measurement as a kind of side information. In  quantum signal detection theory, the difference between "before measurement" and "after measurement" on the knowledge of key which control the quantum measurement is essential. 
In the classical channel, there may be no difference for the order of the knowledge of key. That is,
\begin{equation}
H(X_A|Y_E, K) =0
\end{equation}
From the above result, one may denote
\begin{equation}
R^*_g = \max_{p(X_A)}[I(X_A; Y_B) - I(X_A; Y_E|K)]
\end{equation}
By choosing a rate below the key rate, one is able to force the Eve's information to be zero as a consequence of the well known coding theorem. 

\section{Basis for concrete security analysis}
\subsection{Quantum measurements}
Let us denote here several kinds of quantum processing which can be used by Eve. We can divide broadly into two categories:
Individual  and Collective measurements, respectively. Both of them are well defined by quantum detection theory based on Helstrom-Holevo-Yuen formalism. The former is that
Eve prepares the probe/interaction to each qubit of the quantum signal sequence individually and identically, and processes the resulting information independently from one qubit to the other.
Then she employs classical joint processing.
On the other hand, the qubits may be correlated through the running key $K'$. So one can take a joint attack which requires correlated qubit measurement so called collective measurement. In the each category, one has two kinds of measurement method.
That is, if Eve tries to measure directly the information bits $\{0,1\}$, or $M$ values to $M$-ary modulation scheme, then it is called {\it direct observable attack}. On the other hand, if Eve tries to use some information on the structure of modulation and so on, and measure the indirect observable which can be derived from the structure, then it is called {\it indirect observable attack}.

In addition, there is another quantum measurement scheme so called unambiguous measurement. This has no advantage in communication theory, but it is applicable to cryptography [13,14]. However, here we have the following property[15].\\
{\bf Theorem 3}\\
{\it The lower bound of inconclusive probability in unambiguous measurement is given by the  quantum optimum solution in quantum detection theory for the same state ensemble.}\\

As a result, we can evaluate the limitation of unambiguous measurement attack by quantum detection theory.

\subsection{Randomization in physical layer}
In general, one uses additional several randomizaions in the conventional stream cipher. 
The first demonstration of direct encryption as Y-00 consists of LFSR as PRNG and $M$-ary modulator[8,9]. In addition,  new  randomizaions for  Y-00 are employed, which involves very different concept. We introduce more concrete one here.
The conventional randomization is of data, and it is designed by a mathematical relation. However, Y-00 is designed by both  mathematical relation and physical layer as modulation scheme. So one can take new  parameters  for randomization. For example, modulation, synchronization and so on which are parameters of physical layer in the communication protocol. A new function of this randomization in physical layer is to stimulate making random error for Eve's measurement. This type of randomization provides very good security performance even the system is classical and noiseless. The system consisting of only such a new randomization is called classical Y-00. 
Although the classical Y-00 is stronger than conventional stream cipher, it cannot have information-theoretic security or unconditional security, because it will be insecure under specific side information. For our purpose, one needs irreversible noise effect based on quantum mechanics. \\
{\bf Proposition 1}: \\
{\it The sufficient condition for information-theoretic security on known/chosen plaintext attack is that irreversible error for $Y_E$ is induced by physical randomization with quantum effect.}\\
Proof: \\
In the known plaintext attack, when $Y_E$ does not involve error(no error by measurement), we have $Y_E=Y_B$. 
So $H(K|Y_E, R_P, X)=0$.
The condition for nonzero of key equivocation is that the key is not determined uniquely by $Y_E, R_P, X$. To realize such a situation, the key equivocation should be nonzero even when Eve is allowed to get the randomization schemes used by users after measurement. If error is recovered by side information, the key equivocation is zero. If error is irreversible, then the key is not uniquely assigned by any side information. Since the classical noise is, in principle, removable, we need really irreversible quantum effect which corresponds to the projection postulate.

In conventional system, it is  assumed that error in measurement process is zero, and error by lack of information for mathematical relation(mathematical randomization and so on) can be recovered by side information. In quantum system, 
the randomization and quantum noise effect help each other to make a secure communication scheme by inducing irreversible error in the measurement. 
We can show a concrete example here so called {\it  Overlap Selection Keying}:{\bf  OSK} [16] which is defined as making Eve's density operators $\rho_1=\rho_0$ in the case of direct observable, and  $\rho_{up}=\rho_{down}$ in the case of indirect observable as shown in later sections. So Eve cannot get any information by her measurement. But the realization methods of OSK are adapted to Eve's attack.
OSK which produces the identical density operators for  Eve's any kind of measurement is indeed physical randomization. Thus, the effect of quantum noise is diffused by the randomization.
On the other hand, 
we can introduce an error inducement randomization({\bf EIR}) by physical process also. 
The EIR means to induce error in the discrimination for the signals based on {\it physical force} in communication scheme.
The typical one is to break the synchronization rule between Alice and Eve by changing the synchronization rule between Alice and Bob.
Even if the synchronization errors are few slots, then the error is induced as measurement error.
Such errors are not recovered by the knowledge of the EIR. 
We will show some examples of the effect of  OSK and  EIR in the  section 4.

\subsection{Design of number of basis in M-ary modulation}
One of features in Y-00 is that one can  use conventional laser light as the transmitter which has mesoscopic energy.
Since Eve can attack at the transmitter side, we have to design the number of basis to keep the non-orthogonality among quantum states at the transmitter. In the phase modulation scheme, the coherent states are described by positions on circle in the phase space representation. The radius corresponds to the amplitude or average photon number per pulse at the transmitter.The positions on the circle correspond to phase information of the light wave. If the number of basis is $M$, then the signal distance between  neighbor states is about $\frac{2\pi |\alpha|}{2M}$.
In the practical sense, we can design the number of basis which satisfies 
\begin{equation}
P_e(i+1, i) = \frac{1}{2} - \frac{1}{\sqrt{2\pi}}\int_{0}^{t_0}\exp(-t^2/2) dt
=0.45 \sim 0.5
\end{equation}
where $t_0= \frac{\pi |\alpha|}{2M}$. This corresponds to the error probability between neighbor states. As a result, we have to use at least $M=10^3$ for $<n> =|\alpha|^2=10^4$, and 
$M=10^4$ for $<n> =10^5$.

\section{Quantum individual measurement attack}
In the case of direct encryption, two attacks are important. One is ciphertext only attack on data and key. Other is known/chosen  plaintext attack. We will discuss on both attacks to the system implemented by Northwestern university group so called $\alpha\eta$ scheme [8,9] based on $M$-ary phase shift keying by coherent states. In general, the phase spaces for Alice-Bob, and Alice-Eve are different, because the phase used in the communication means relative phase between transmitter light and reference light at receiver. 
Although the phase spaces are not same in general, we first assume that the phase spaces of three parties are same. 

\subsection{Ciphertext only attack}
Now the seed key in the system has the relation $H(X) >> H(K)$.
In the conventional theory, we have always $H(X_A|Y_E, R_M) \le H(K)$ for ciphertext only attack on data. 
But this is not true in the case of Y-00. In the following we will show some examples.

\subsubsection{ Direct observable attack}
When Eve employs direct measurement on data or on key, the ultimate error probability is given by Eq(3) based on quantum detection theory for each bit slot. When we design the system by appropriate photon number and $M$ based on Eq(12), the error probability is almost $\frac{1}{2}$. In addition, by using OSK, it becomes exactly $\frac{1}{2}$ because of $\rho_1 =\rho_0$. These fact mean that Eve cannot get any information on data by her measurement.

For the property of {\it reuse of the key}, one can check the processing $Y(1)\oplus Y(2)=X(1)\oplus X(2)$, where $Y(1)=X(1)\oplus K', Y(2)=X(2)\oplus K'$, and where $K'$ is PRN. Thus, in general, the key disappears in the stream cipher, while the one time pad has $Y(1)\oplus Y(2)=X(1)\oplus X(2)\oplus K_1 \oplus K_2$. 
$K_1$ and $K_2$ are true random number. 
However, in Y-00, the measured values of Eve have $Y_E \ne Y(1 {\rm or} 2)$, but $Y_E=Y \oplus \bf{e}$, where $\bf{e}$ is error vector and it is completely random  by quantum noise effect. So we have  $Y_E(1)\oplus Y_E(2)=X(1) \oplus X(2)\oplus \bf{e}_1 \oplus \bf{e}_2$. Since $\bf{e}_1 \oplus \bf{e}_2$ is completely random when $\bar{P}_e = 1/2$, the performance is equivalent to the one time pad.

\subsubsection{ Indirect observable attack}
Since the modulator has definite structure, Eve can use an information on such a scheme. Indeed, the randomly selected phase shift keying used in  $M$-ary cipher scheme as Y-00 is taken to be
\begin{equation}
l_i=x_i\oplus \tilde{k}_i
\end{equation}
on the phase space, where $l_i$ is one of two regions separated by appropriate axis on the phase space.
If the fundamental axis is horizontal, $l_0$ is upper plain, $l_1$ is down plain, $x_i$ is data bit. $\tilde{k}_i$ is 0 for even number and 1 for odd number in the running key of $M$-ary assignment [15,16]. However, we should denote that $\tilde{k}_i$ is the result of the mapping from the running key of decimal number:$k_i = (1 \sim M)$.
\begin{equation}
K'_j =(k_1, k_2, k_3, \dots) \mapsto \tilde{K}_j =(\tilde{k}_{1}, \tilde{k}_{2}, \tilde{k}_{3}, \dots), \quad \{j=1 \sim 2^{|K|}\}
\end{equation}
where $ \tilde{k}_{i}= even$ or $odd$.
For example, 
$(l_i=up,\quad \tilde{k}_i=even) \longrightarrow x=1$, 
$(up,\quad odd) \longrightarrow x=0$, 
$(down, \quad even) \longrightarrow x=0$, 
$(down,\quad odd )\longrightarrow x=1$. 
Let us define the sequences of numbers $l$, $x$, $\tilde{k}$ as follows: $L=(l_1, l_2, l_3, \dots), X=(x_1, x_2, x_3, \dots), 
\tilde{K}=(\tilde{k}_{1}$. 
Here $K$, and $N$  are an initial key with length $|K|$, and length $|N|$ of pseudo random number, respectively.
The essential point of the attack is to measure indirect observable $L$. However, since the observable does not contain the information of the data bit, Eve is asked to use the brute force attack for key to find a correct sequence of the data.
Here we can define that  $\cal{R_A}$ is a set of data  sequence with the length of $|N|$.
Alice sends a sequence  $R_T$ in $\cal{R_A}$, and it is coded based on Eq(13)[8,9]. 
$\tilde{K}_j$ corresponds to pseudo random number sequence which  has at most a number of possibilities of $2^{|K|}$ and the length $|N|$.
Let  $R_T$, $L_T$, $\tilde{K}_T$ be true sequences used and defined on the phase space for Alice and Bob.
She tries to assign  all kind of $\tilde{K}_j$ to her measured sequence $L_m$ of $L$. So she gets 
a set $\cal{R}_{\rm{E}}$ based on $l_i=x_i \oplus \tilde{k}_i$. If $L_m$ is error free, then it is guaranteed that one of $\cal{R}_{\rm{E}}$ is the data bits sequence. At this stage, we have $H(X|Y_E)=0$ based on the brute force attack. 
Here, if there is one bit error in $L_m$ by some reasons, then Eve has $L_T\oplus \bf{e}$, where ${\bf{e}}=(0,0,1,0,0, \dots)$ is error sequence. The position of the error is unknown and uniformly distributed. When Eve applies $\tilde{K}_j$ to $L_T\oplus \bf{e}$, then  it is not guaranteed that the true $R_T$ exists in $\cal{R}_{\rm{E}}$.
She has to try $2^{2|K|}$ greater than the initial one. If there are many error, then it becomes 
$\sim 2^{2^{|K|}}$. So one may obtain $H(X|Y_E) > H(K)$.

Indeed, let us show that the error of the measurement for $l_i$ is unavoidable. The density operators of signal sets for up and down measurement are 
\begin{eqnarray}
\rho_{up}&=&\sum_{up} \frac{1}{M}|\alpha_{i}\rangle \langle \alpha_{j}|, \\
\rho_{down}&=&\sum_{down} \frac{1}{M}|\alpha_{j}\rangle \langle \alpha_{j}|
\end{eqnarray}
It is easy to show the quantum  limit, which is the most rigorous lower bound of error probability for this signal[17,18]. When the coherent state is mesoscopic$<n>\sim 10000$ and one thousand of $M$ based on Eq(12), the error is several percents: $P_e \sim 0.01$. This means that the number of error bits is $P_e \times 2^{|K|} \gg 1$ which is enough to prevent this type of attack with the brute force search. That is, we have $H(X|Y_E)>H(K)$. For the performance of reuse of key, we have $L_1\oplus L_2=X_1\oplus X_2 \oplus \bf{e}_1 \oplus \bf{e}_2$.
If the number of error is small, the security may depend on $H(X)$. However, the error probability can be increased by arranging the signal state assignment. 

Furthermore, again we can employ OSK. 
Since the best way for indirect observable attack is to divide the phase space into two regions like up plain and down plain,  Alice can employs {\bf OSK} which provides $\rho_{up}=\rho_{down}$. 
As a result, $\bf{e}_1 \oplus \bf{e}_2$ of $L$ is completely random, and it has no information on transmitting states. Such an OSK is made by changing the fundamental axis of the phase space of Alice-Bob based on a part of running key, for example $0-\pi$, and $\pi-0$. Thus, if one uses randomizations as OSK, then the scheme is secure to the ciphertext only attack on data  even in the noiseless.
If users require a security against only the ciphertext only attack, then "quantum effect" in Y-00 may be not essential, though it is helpful. That is, {\it the classical Y-00 with several randomizations is sufficient enough for ciphertext only attack}.

\subsection{Known/chosen plaintext attack}
Since the ciphertext only attack on key is absorbed into the known/chosen plaintext attack, we skip it.
In the conventional cryptography, the known/chosen plaintext attack means that Eve can get many pairs between known data bit sequence and corresponding ciphertext. In Y-00, a ciphertext corresponds to a sequence of quantum states. 
Thus, in this case, Eve knows several set of plaintext and corresponding quantum state sequences.
So Eve has an additional problem such as quantum state identification which is done by quantum measurement.
Eve will have two methods which utilize such an additional knowledge for improving her attack. One is to reduce the number of selection of quantum measurement, and other is to use it after measurement which corresponds to Lo-Ko attack[19].

\subsubsection{ Conventional method}
Let us assume that Eve can insert known sequences as data. So the output quantum state sequence of the $M$-ary modulator correspond  to  running key information. In the case of individual attack, Eve can try to discriminate $M$  coherent states by quantum optimum receiver. It is well known that one can find a seed key or solve next bit prediction for LFSR if one knows exactly  bit sequence of 2$|K|$ as the running key. However, in Y-00, the error in the measurement is unavoidable.
Although there is no general theory of lower bound of algorithm for finding key based on running key information with error, in the limit, we may have the fact that the error probability is given by Eq(4), which converges 1 with respect to large number of $M$. 
As a result, by appropriate number of $M$ and photon number, 
we have $H(K|Y_E, X) > 0$ which means information-theoretic security.

As an example, we here give more concrete scheme. Y-00 has the following structure. The output of LFSR is divided by $log M$ block, and the bit sequence of the length $log M$ is changed into the number $k_i \in {\cal M}$ which corresponds to the numbering of basis. So the output state is a coherent state with the same numbering. Here the number is assigned as regular order on the circle of the phase space.
Let us assume that Eve knows plaintext with length of $2|K|$, say $X_a$.
By  heterodyne receiver, Eve measures the quadrature amplitude $x_c$ and $x_s$ putting known plaintext:$X_a$ to decide which basis is used. The errors of measured data of Eve are induced  mainly for the neighbor quantum states. That is, when the measured number is 5, then it has possibility of 4 or 6 as true number. Since the number of  slot is $2|K|/log M$, the measured sequence will be, at least,  one of the number of sequences $2^{2|K|/log M}$. This may correspond to 1 or 2 bit random error per $log M$ bit block in the bit sequence of the output of LFSR.  
If the error occurs among several neighbors, then Eve suffers more error in corresponding bit sequence.
Here, since the error bits of measured data in bit sequence are  $\sim (2|K|/log M)=T$ bits,  Eve has to launch $W=2^T$ times  the next bit prediction algorithm. By design of $M$ and $<n>$ based on Eq(12), 
we have $W>>2^{|K|}$.\\

\subsubsection{ Lo-Ko attack}
Here let us assume that Eve will divide the light wave conveying  known plaintext $X_a$ into $W$  beams by beam splitter, then she will prepare $W$ receivers with different key and measures each beam by each receiver. Since Eve knows the plaintext, the receiver which outputs the same plaintext is correct one. Such a method is called Lo-Ko attack[19].
Let us assume that the amplitude attenuation parameter of channel between Alice and Bob is $\kappa=1/(t+1)$. The amplitude of Bob is given by $\alpha/(t+1)$. Eve makes $(t+1)$ copies by means of division of the output light from Alice by $(t+1)$ beam splitters.
\begin{eqnarray}
|\Psi \rangle &=&|\frac{\alpha_i}{t+1} \rangle |\frac{\alpha_j}{t+1} \rangle |\frac{\alpha_k}{t+1} \rangle \dots \nonumber \\
|\Psi \rangle &=&|\frac{\alpha_i}{t+1} \rangle |\frac{\alpha_j}{t+1} \rangle |\frac{\alpha_k}{t+1} \rangle \dots \nonumber \\
|\Psi \rangle &=&|\frac{\alpha_i}{t+1} \rangle |\frac{\alpha_j}{t+1} \rangle |\frac{\alpha_k}{t+1} \rangle \dots \\
\vdots \nonumber
\end{eqnarray}
Then the first sequence is sent to Bob by a lossless channel.
Bob cannot notice the existence of Eve, because his coherent state sequence is exactly same as that of regular communication.
Bob will employ appropriate receiver with assigned key in which the performance will be error free(actually it is not error free, and they use error correcting code). 
Eve has to try all kinds of receiver. 
So she needs exactly $W=2^{|K|}$ copies.
Since she knows plaintext as input data, she can compare her measurement result with the plaintext. The receiver which showed the same result with plaintext is Bob's receiver, and she can know the key. 
However, the amplitude of each coherent state is scale down by factor $1/(t+1)$. In this case, in order to get $W$ copies, the requirement for loss between Alice and Bob is $1/t =1/W$. 
It is easy to show that Lo-Ko attack does not work for practical situations, for example, if the communication length by conventional optical fiber is 100 km, then $t=100 << W=2^{|K|}$. This means that Eve cannot get enough copies which is able to decide the key.\\

\subsubsection{ Modified Lo-Ko attack}
Let us assume that the loss is realistic such as 
$\kappa >> 2^{-|K|}$. By first splitter, Eve makes 
\begin{equation}
|{\Psi}' \rangle =|\kappa \alpha_i \rangle |\kappa \alpha_j \rangle |\kappa \alpha_k  \rangle \dots 
\end{equation}
This is sent to Bob by a lossless channel.
Then she makes $2^{|K|}$ copies as follows:
\begin{eqnarray}
|\Psi \rangle &=&|(1-\kappa)\frac{\alpha_i}{W}\rangle |(1-\kappa)\frac{\alpha_j}{W}\rangle |(1-\kappa)\frac{\alpha_k}{W} \rangle \dots \nonumber \\
|\Psi \rangle &=&|(1-\kappa)\frac{\alpha_i}{W}\rangle |(1-\kappa)\frac{\alpha_j}{W}\rangle |(1-\kappa)\frac{\alpha_k}{W} \rangle \dots \nonumber \\
|\Psi \rangle &=&|(1-\kappa)\frac{\alpha_i}{W}\rangle |(1-\kappa)\frac{\alpha_j}{W}\rangle |(1-\kappa)\frac{\alpha_k}{W} \rangle
\dots  \\
\vdots \nonumber
\end{eqnarray}
However, in this case, the amplitude of Eve's quantum state sequences is smaller than that of Bob. So even if Eve has 
$W$ copies, since the performance of her measurement is less than that of Bob, her results involve unavoidable error.
As a result, the knowledge of plaintext cannot help to decide Bob's receiver.\\

\subsubsection{ Indirect observable attack and others}
Let us discuss known/chosen plaintext attack based on indirect observable. Eve knows certain relation of the modulator like $l_i=x_i\oplus \tilde{k}_i$. 
Assume that Eve has several known plaintexts. For measured data of $l_i$, she applies the known plaintext. So she can get the sequence of $\tilde{k}_i$ based on the above relation. However, the running key sequence cannot be determined by the sequence of $\tilde{k}_i$, because Eq(14) is not one to one correspondence.
In addition, as we mentioned in the case of ciphertext only attack, the number of error slots on the measurement of $l_i$ are not so small even the error probability for $l_i$ is small, when the system is well designed based on Eq(12). 

Then, let us apply unambiguous measurement attack.
On the prediction of the key sequence based on known plaintext, in the conventional stream cipher by LFSR, the requirement of length of known plaintext is about the key length. 
In order to realize the unambiguous measurement attack, we need collective measurement. The collective measurement means that a discrimination problem of many quantum state sequences is treated such that a quantum state sequence can be regarded as one pure state in the tensor product Hilbert space of time slot modes, and the design of the detection operator:$\Pi^{N}$ as POVM is formulated on the extended space.
Since Eve knows structure of PRNG and modulation scheme, she can apply the unambiguous measurement:$\Pi_{un}^{N}$ for $2^{|K|}$ quantum state sequences to one copy of the transmitting sequence on the extended space. When the length of key is enough long, the success probability is given based on the theorem 3 as follows:
\begin{equation}
P_d \sim 2^{-2|K|}
\end{equation}
where we use $|K| > 100$. So with this probability, Eve will get an exact running key sequence. However, in principle, Eve cannot get a situation which the success probability is 1.\\

\subsection{Randomization for long distance communication}
In order to realize secure long distance communication, one can use more strong light power at the transmitter. In such a case, the error may be very small in Eve's any measurements.
However, we can cope with such a situation by introducing several kinds of physical randomization pioneered by Yuen.

To prevent the several attacks, 
we can introduce the randomization for numbering to assign the basis state which is one example of DSR. As a result, the error position in the bit sequence of running key can be uniformly diffused into the total length of the running key.
On the other hand, in the above sections, we assumed that Eve can synchronize the phase spaces between Alice-Bob, and Alice-Eve. In general situation, it is difficult. Even it can do at the first step, legitimate users can break easily the synchronization by randomization or based on seed key advantage.
Let us denote more detail nature. Since Y-00 is a physical cryptography, we should clarify the physical property. In general, phase spaces of Alice-Bob and Alice-Eve are not  same. The phase space is formed by the relative phase based on local phases of Bob and Eve. For example, quadrature amplitudes are  $\{x_c=A\cos(\phi_S-\phi_{L(Bob)}), x_s=A\sin(\phi_S-\phi_{L(Bob)})\}$, 
$\{x_c=A\cos(\phi_S-\phi_{L(Eve)}), x_s=A\sin(\phi_S-\phi_{L(Eve)})\}$. The Eq(13) is defined for the phase space of Alice-Bob. In general Eve does not know the correct phase space. 
Thus it is easy to control the fundamental axis of the phase space to prevent Eve's locking.
This fact is one of characters of physical cryptography. So Eve never understands what is the  axis decided by herself. This is one of EIR.
As a result, Eve's data involves many errors, even when the measurement is noiseless. This is a concrete example why Y-00 is stronger than that of the conventional scheme even the system is classical one. When such randomizations and quantum noise effect are used, the quantum noise effect is diffused. As a result, the quantum noise effect is enhanced.

\section{Key generation}
In the key generation, since data is true random number, Eve's strategy is mainly to apply direct observable attack. In addition, the attack to get the running key (or seed key) based on direct observable of ciphertext is hopeless for Eve as mentioned in the previous section. So Eve has to try to get information of random number directly.

\subsection{Key rate based on entropy evaluation}
In this section, we would like to obtain certain example of the key rate in the framework of Yuen's general discussions.
From lemma 1 in the reference [4] and Holevo bound theorem, we have
\begin{equation}
I(X_A(n), Y_E(n)|K)/n < I(X_A(n), Y_E(n))/n + H(K)/n < S(\rho_T) - \sum p(x) S(\rho_{x}) + H(K)/n
\end{equation}
where $I(X_A(n), Y_E(n))$ refers to the mutual information between $n$-bit sequences, $S(\rho)$ is von Neumann entropy, and where
\begin{equation}
\rho_T = p(0)\rho_0 +p(1)\rho_1
\end{equation}
$\rho_0$ and $\rho_1$ are given by Eq(2).
Let $\epsilon$ be cp-map between Alice and Bob, and let us assume that the channel between Alice and Eve is ideal channel(cp-map is identity map), because Eve can make her measurement at the close Alice. 
As a result, we can read as follows:
\begin{equation}
r_{g(n)} \ge \max_{{p}(i)} [\{S(\epsilon ({\rho}^{B}_T)) - \sum {p}(i)S(\epsilon ({\rho}^{B}_i)\} - 
\{S({\rho}^{E}_T) - \sum {p}(i)S({\rho}^{E}_i) + H(K)/n \}]
\end{equation}
When the cp-map corresponds to ideal linear attenuation with attenuation parameter $\kappa$, Bob's density operator is
${\rho}^{B}_0 = |\kappa\alpha\rangle\langle \kappa\alpha|$, 
${\rho}^{B}_1 = |-\kappa\alpha\rangle\langle -\kappa\alpha|$, 
and that of Eve is Eq(2) for the direct observable.
When $n$ is enough large, we can calculate it for the concrete scheme described above. For example, we assume that $<n>=1000$,  $M=100$, $\kappa=1$, and the data rate is 1 Gbps. Then the secure key bits are $10^7$. However, with respect to $\kappa$, the secure key bits decrease. It has only 6 dB energy loss advantage to keep the secure non-zero key bits.

When we apply OSK for direct observable, we 
have
\begin{equation}
I(X_A;Y_E)=I(X_A;Y_E|K) = 0
\end{equation}
As a result, key rate is equal to the Holevo capacity of the channel between Alice and Bob. Thus we can see the great advantage of the diffusion of quantum noise by randomizatons.

\subsection{Combined attack to quantum and classical channels}
Let us assume that the generated key is used as symmetric key in classical channel communication like one time pad, and Eve can get perfect ciphertext on the classical channel.

If Eve takes direct observable on the quantum channel,then Alice and Bob can share the key bits as mentioned above. Even Eve gets all ciphertext on the classical channel, the key in the ciphertext has no correlation between Eve's prediction and real key for one time pad, because of privacy amplification.

On the other hand, we can consider the indirect observable in this case. Nishioka et al discussed on the following attack[20,21].
Eve measures $l_i$, and she does not assign $\tilde{K}_j$, $j \in 2^{|K|}$ at the first stage for quantum communication used Y-00. Then she get the data $c_i$ on the classical channel.
In this model,  one has the relations as follows:
\begin{eqnarray}
c_i&=&x_i \oplus r_i \\
c_i \oplus l_i &=& x_i \oplus \tilde{k}_i
\end{eqnarray}
where  $c_i$ is ciphertext, $x_i$ is plaintext.
The random number $r_i \in R$ as the key disappears.
Since Eve does not know $\tilde{k}_i$, she has to  assign to all the different running keys in which 
the number of possibilities is $2^{|K|}$ . 
If Eve can get the data $l_i$ and $c_i$ without error, then it corresponds to conventional noiseless stream cipher with exponential search. However, by the effect of quantum noise and randomizations like OSK and DSR, Eve's data of $l_i$ are completely random. So such an attack has no meaning.
We emphasize that the original experiment[8] of Y-00 did not claim this kind of protocol. So this kind of attack is a fiction.

\section{Intensity modulation scheme and its experiment}
We would like to devise an attractive technology based on quantum communication theory which is applicable to the real world. 
It is well known that intensity modulation based optical communication is widely used in the conventional optical fiber network system. Of course Y-00 is applicable to it. However, in general, it is difficult to increase the number of $M$ in the intensity modulation scheme. So we need appropriate randomizations. 
In the reference [16], we proposed Y-00 based on intensity modulation and direct detection(IMDD) scheme with appropriate randomization so called OSK which was reviewed in the section 3.
Let us give again a brief explanation of the scheme.
The maximum amplitude of the transmitter is fixed as $\alpha_{max}=\alpha_{2M}$.
We divide it into 2$M$. So we have $M$ sets of basis state
$\{(A_1,A_2),(B_1,B_2), \dots \}$. 
The total set of basis states becomes
as shown in Fig.1. 
\begin{figure}
 \begin{center}
 \begin{tabular}{c}
 \includegraphics{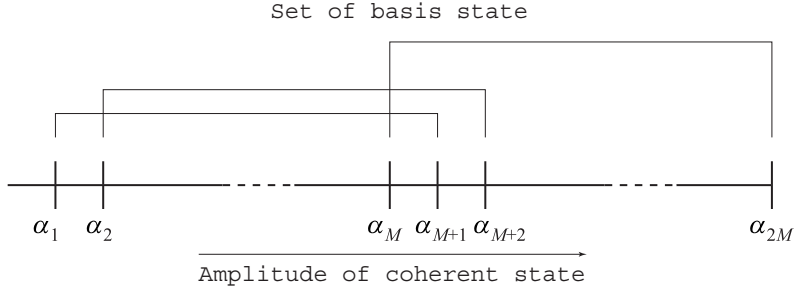}
 \end{tabular}
 \end{center}
\caption{M-ary scheme based on amplitude(intensity)}
\label{figure1} 
 \end{figure}
The error probabilities of Bob and Eve in this case can be derived from our papers [17,18].
Here let us employ OSK for data bit sequence and also for the numbering of the basis $M$ related up and down on the axis for amplitude.
Each set of basis state is
used for $\{1, 0\}$, and $\{0, 1\}$, depending on sub-running key.
\begin{eqnarray}
Set\quad A_1 : 0 &\rightarrow& |\alpha_{(1)}\rangle, 
\quad 1 \rightarrow |\alpha_{(M+1)}\rangle\\
Set\quad A_2 : 0 &\rightarrow& |\alpha_{(M+1)}\rangle, 
\quad 1 \rightarrow|\alpha_{(1)}\rangle
\end{eqnarray}
So the density operators of 1 and 0 for Eve are 
$\rho_1=\rho_0$.
Furthermore, by changing the numbering of the basis from up to down by sub-running key, the density operators for Eve become 
$\rho_{up}=\rho_{down}$.

In order to verify the advantage distillation of Y-00 based on the intensity modulation scheme, we show  experimental result on error performances of receivers of Bob and Eve. The system consists of the conventional laser diode and photo diode which work under the 5 Mbps and room temperature. The number of coherent states are 4, but the amplitude difference is very small.
We assume that the technology level of Bob and Eve are the same one. Figure 2 shows the error probabilities  of Bob  who knows key and of Eve who does not know key, respectively, when the legitimate users do not use the randomizations. Y-00 can work under the distance corresponding to the difference of the error probability. On the other hand, when we employ the OSK, the error probability of Eve is 1/2. So the communication distance is limited only by the error performance of Bob. The effect of the OSK is very clear in this case.

\section{Conclusions}
In this paper, it has been discussed that Y-00 based on IMDD realizes a scheme with provable security in the sense of information theoretic security against at least several proposed attacks, which is not realized by  conventional cryptography.
If one requires the security for ciphertext only attack, then one does not need "quantum", but classical Y-00 is enough.
Thus, it is clear that recent criticisms [19, 20] on Y-00 have no meaning in the sense of cryptography.

\section*{Acknowledgment}
OH is grateful to H.P.Yuen, P.Kumar and many colleagues of Northwestern University for discussions.


\begin{figure}[b]
 \begin{center}
 \includegraphics[scale=0.7]{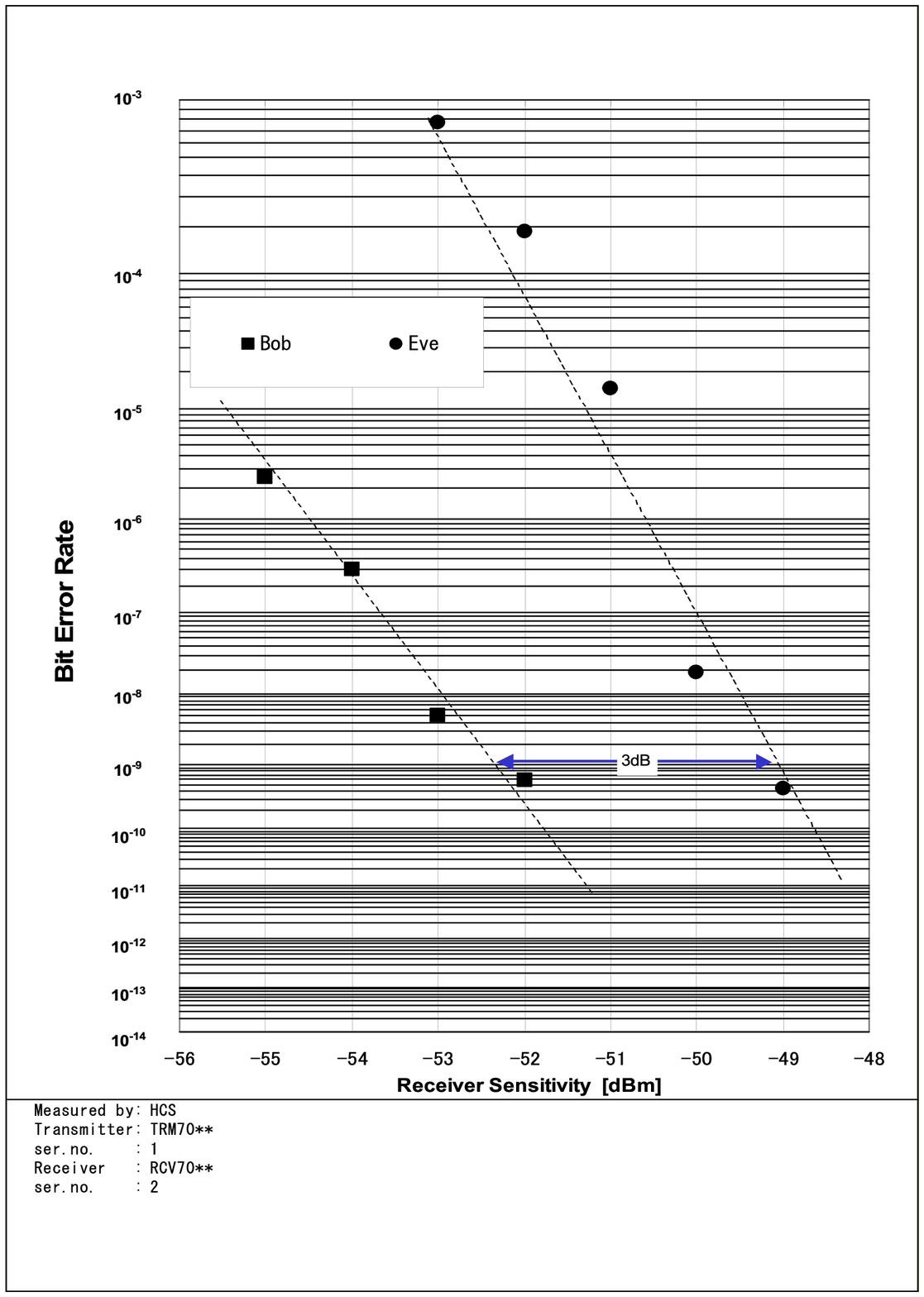}
 \end{center}
\caption{Error probabilities of Eve and Bob}
\label{figure2} 
 \end{figure}
\end{document}